\documentclass{ijmpi_authors}

\usepackage{lipsum}
\usepackage{fullpage,graphicx}
\usepackage{tikz}
\usepackage{mathtools,amsfonts,amsthm,multirow}
\usepackage{algorithmic,comment,url,enumerate}
\usepackage{amsmath} 
\usepackage{tikz}
\usepackage{relsize}
\usepackage{float}
\usepackage{framed}
\usepackage{hyperref}
\usepackage{enumitem}
\usepackage{xcolor}
\usepackage{caption}
\usepackage{subcaption}
\usepackage[ruled,vlined]{algorithm2e}

\DeclareMathOperator*{\argmin}{arg\,min}
\newcommand{\E}{\mathbb{E}}

\hypersetup{
    colorlinks=true,
    linkcolor=blue,
    filecolor=magenta,      
    urlcolor=cyan,
    citecolor=blue
}

\begin{document}

\twocolumn[
\title{Almost-Nash Sequential Bargaining}

\author{Dharan}{Gokul}{s}
\author{Guru}{Hunter}{s}
\author{Sun}{Michael}{s}

\affiliation{s}{Stanford University}

\maketitle

\begin{abstract}
In a 2017 paper, later presented at the Web and Internet Economics conference, titled ``Sequential Deliberation for Social Choice" \cite{fain2017sequential}, the authors propose a mechanism in which a series of agents $N$, are tasked to negotiate over a set of decisions $S$. Building on assumptions of Nash Bargaining and assuming the decision space follows the median graph, the authors constructed a robust algorithm which approximates the decision which minimizes the social cost to the entire population. In this paper, we give a brief overview of the background theory which this paper builds upon from foundational work from Nash, and social choice results which hold true in Condorcet mechanisms. Following this analysis, we consider the stability of the results in the paper with different deviations from Nash equilibrium. These deviations could be pessimal, in the context of unequal bargaining power (say in a labor market) or constructive, as in the context of opinion dynamics. Our analysis is observatory, in the context of simulations, and we hope to formalize the results of these simulations to get an understanding of more general properties in spaces beyond our simulation.
\end{abstract}

]

\section{Literature Review}

\subsection{Introduction}

Consider an environment $\mathcal{E} = (N, S)$ in which there is a population of participants (individual decision making agents), $N$, and a set $S$, corresponding to the set of decisions / outcomes available to them. The goal of the mechanism designer is to pick a socially desirable outcome, $o \in S$, which leads to an equilibrium corresponding to consensus over the agents.

Typically in these kinds of settings, one could imagine polling each agent for their preferences over $S$ and aggregating these preferences to pick such an alternative. Unfortunately, this "enumeration" technique is not always feasible as the space $S$ may not even be completely known to the mechanism designer. Additionally, as the size and complexity of these spaces grow, simply working with enumerations is tough \cite{lang_xia_moulin_2016}. 

The motivation of the mechanism designer could very well be choosing an option which minimizes the \textit{social cost} of an alternative. Here we adapt much of the structure proposed in \cite{fain2017sequential}. Assume each agent $i \in N$ has a "bliss point", $p_i$, which is their most preferred element in $S$. If the designer is minimizing the social cost by picking an alternative, $a$, they are picking the quantity which minimizes

$$
a^* := \argmin_{a \in S} \sum_{u \in N} d(p_u, a) := \argmin_{a \in S}SC(a)
$$
where $d(x, y)$ to refer to the disutility of an alternative $y$ to another alternative $x$. Notice that the notion/function representing $d(\cdot, \cdot)$ is common to the agents. For any $x, y, z \in S$, then, it is clear that we would want $d$ to obey $d(x, z) \leq d(x, y) + d(y, z)$, to codify rationality amongst the agents.

So, in such a model, the mechanism designer would want to choose an outcome $\hat{a}$ which has $SC(\hat{a}) \approx SC(a^*)$. We adopt the convention of the authors by utilizing distortion as a metric to analyze this closeness:

$$
\mathcal{D}(\hat{a}) := \text{Distortion}(\hat{a}) = \frac{SC(\hat{a})}{SC(a^*)}
$$
and since $a^*$ is a minimizer of $SC(\cdot)$, $\forall a \in S$, we have $\mathcal{D}(a) \geq 1$. 

\subsection{Introducing the Model}

Rather than restricting agents to pick and order alternatives on the ballot, we investigate the implications of the deliberative forum proposed in \cite{fain2017sequential}. The concept of deliberation is that individuals can communicate with one another in a negotiation process, in which they generate new alternatives. For example, rather than some students ranking a set of particular grading schemes and aggregating their choices, we consider the model where the students actually negotiate and come up with alternatives, relative to some fixed alternative. In this analysis, only pairwise deliberation is considered; that is, not more than two individuals are deliberating simultaneously. 

\subsubsection{A Brief  History: Nash Bargaining}

More formally, deliberation closely resembles the canonical model for two-player negotiation \cite{Narahari_2012}, originally formulated by John Nash \cite{nash1950bargaining}. Two-player bargaining is formulated as a tuple ($\mathcal{F}, t$), where a pair of players attempt to pick an outcome from a feasible set $\mathcal{F}$, and if they are unable to come to a conclusion, then the point which they default to is $t$ -- or the threat/disagreement point. Nash proved that a game in which $\mathcal{F}$ is a convex set with the requirements of:
\begin{itemize}
    \item Pareto optimality: The outcome is such that neither agent can improve without hurting the other.
    \item Symmetry between agents $u, v$: Given that $t_u = t_v$ If $f^* = (f^*_1, f^*_2) \in \mathcal{F}$ is selected, then it should be the case that $f^*_1 = f^*_2$
    \item Scale Covariance: An affine transformation of either player's utility function does not change the outcome. 
    \item Independence of Irrelevant Alternatives: Eliminating an alternative should not affect the decision, that is if there is a subset of the feasible set in which the same allocation is available, the agent should stay consistent.
\end{itemize}
corresponds directly to maximizing the Nash product. That is, given a threat $t$, $o^*$ is defined as the $o \in S$ which maximizes
\begin{equation}\label{eq:1}
    (d(p_u, t) - d(p_u, o)) \times (d(p_v, t) - d(p_v, o)).
\end{equation}

A mechanism designer can simulate this process amongst two agents with a repeated game. All that needs to be done is to introduce a decay parameter $\delta^t$, and ask the agents to play a repeated game in which the agents go back and forth making offers -- if an offer is not accepted, the agent is forced to accept the threat at that particular round $t_i$. If the probability of the game continuing diminishes at a rate proportional to $\delta$, such a game has a sub-game perfect Nash equilibrium corresponding directly to the product, assuming agents are sufficiently patient.

We note that as the requirement over these axioms dissolves, this game may have many other equilibrium. A perfect example is in the case where $\mathcal{F} = [t, z]$, (ex. courtesy of the Wikipedia article). Consider the strategy/allocation of $(x_1, x_2)$ such that $x_1 + x_2 = z$. Here, if either player attempts to increase their consumption, this is outside the feasible solutions, and clearly decreasing can never help either. This realization motivated us to consider settings in which perfect play was not guaranteed, or other external forces (social dynamics) may incentivize players to play off the standard equilibrium path. Numerous behavioral studies, for example \cite{schellenberg1990solving}, indicate that the assumptions of the model were not always satisfied, and bargaining of individuals did not necessarily correspond to the Nash model. 

In fact, there are other axioms and solution concepts for cooperative bargaining which have risen to cater different needs and assumptions. Loosening the constraint of independent and irrelevant alternatives (IIA) leads to the Kalai-Smorodinsky bargaining solution \cite{kalai1975other}. One could imagine a setting in which IIA would not hold; consider players prefer preferences over feasible allocations $p_1: A \succ B \succ C$ and $p_2: C \succ B \succ A$. In a bargaining context with all three allocation points, $B$ seems like a "fair" allocation outcome as both players compromise their top choice to reach their middle preference. However, in a subset allocation of just $\{A, B\}$, we see that $B$ is still available to both players, but is just as "attractive" as $A$ -- so it need not be the case that IIA holds, but in Nash's model it must be the case that we pick the same allocation from both sets, given we do not pick $C$.

Given that this setting also is implicitly a bargaining problem over $n$ agents, one can also consider a simultaneous bargaining model. The most straightforward technique, relative to the prior work discussed, is to split players into coaltions, and find a way of "aggregating" their preferences \cite{chae2001nash}. In this paper, the authors split the players into partitions and add an axiom (representation of homogenous coalition), treating the partition, collectively, as a single agent. Interestingly, the paper shows motivation of the "joint-bargaining paradox", indicating via counterexample that it is possibly unprofitable for a given set of agents to come together and form a larger coalition. At first this may seem entirely surprising, but is similar to the logic which suggests that mergers in a Cournot game may strictly decrease the profit of the agents in the merger, while increasing the profits of the others \cite{salant1983losses}.

\subsubsection{Random Dictatorship: A Reference Model}

Perhaps the most immediate alternative model to consider is that of random dictatorship, as suggested by Fain in a presentation of this paper \cite{Fain_2017}. The algorithm asks the mechanism designer to randomly pick an agent, $i \in N$, and ask the agent their bliss point, $p_i$. The mechanism designer then imposes $p_i$ over all the agents. 

First, see that this model is Pareto efficient -- this follows immediately from the fact that if there were another allocation, then it would strictly hurt agent $i$. A fairly robust claim is that this model gives an expected distortion factor of at most $2$. This follows immediately from the triangle inequality, and is a proof from the authors \cite{fain2017sequential}:
\begin{align*}
    \E_S[SC(x)] &= \frac{1}{n} \sum_{v = 1}^{|N|} \sum_{u = 1}^{|N|} d(p_u, p_v) \\
                &\leq \frac{1}{n} \sum_{v = 1}^{|N|} \sum_{u = 1}^{|N|} d(p_u, a^*) + d(a^*, p_v) = 2SC(a^*).
\end{align*}
The authors also provide a lower bound, which is also of $2$, and finds a tight bound of the method. But intuitively, this algorithm fares poorly when there is a clear compromise which exists between the agents, as it forces an asymmetric choice by the lucky one who was selected. 

\subsubsection{Iterative Compromises \& Sequential Deliberation}

The authors chose to develop a very generalizable and intuitive framework, which built upon many of the ideas of Nash bargaining in a setting with many agents. Having a simple framework was an important consideration, as it allowed for the feasibility of practical implementation. They utilized two-person bargaining as an operation, to iteratively move towards some larger unified alternative. 

At each round $i \in \{0, \cdots, T\}$, the authors randomly choose a pair of agents $u, v \in N$ uniformly, and ask them to debate over some disagreement alternative, $a^t$. They output the alternative selected at the conclusion of their bargaining, and if they are unable to reach a conclusion, the disagreement alternative, $a^t$ is chosen as the outcome. Then, for the next round, the disagreement alternative/threat becomes $a^{t + 1} \leftarrow o^t$, where $o^t$ was the outcome of the agents' deliberation. After $T$ rounds, $a^T$ is output.

Notice that the authors repeatedly are iterating a bargaining game between the agents, and so they assumed that the choice chosen by the agents corresponded directly to the Nash bargaining solution.

\subsection{Analysis of the Model}

In order to measure the convergence to some particular distortion, the authors categorized certain types of alternative spaces, $S$, and analyzed how well the framework held up in these particular settings. All types of spaces assumed the existence of some metric, $d$.

\subsubsection{The Median Graph}

The first model chosen by the authors to represent the decision space was the median graph. Formally, a median graph satisfies the property that any three nodes $u, v, w \in S$ have a unique intersection point amongst the shortest paths of the three possible pairs $\{(u, v), (u, w), (v,w)\}$ of vertices. Some examples of median graphs include hypergraphs, trees, and a line -- some of which we model in our forthcoming simulations. 

Having no prior experience in social choice research, we were particularly interested in the choice of median graphs to model decision spaces, especially as the results in the paper indicated that the decision space played a significant role in the outcomes of the distortions.  A class of preferences are single-peaked \cite{black1948rationale}, if they satisfy:

\begin{itemize}
    \item Each agent has a peak. That is, a most preferred outcome in the set of feasible alternatives.
    \item Each agent prefers outcomes further from their peak less than those closer to their peak. 
\end{itemize}
When preferences satisfy these criteria, the simple truthful mechanism is to select median preferred quantity, by the median voter theorem \cite{black1948rationale}. The Condorcet winner, $w$, when chosen, satisfies the fact that when all agents vote over any election with $w$ and some other $s \in S$, $w$ would win the plurality vote. The Condorcet criterion is satisfied for a particular mechanism if it always picks the Condorcet winner -- in the example regarding the median voter theorem, the theorem more generally states that any voting mechanism satisfying the criterion picks the $a \in S$ "nearest" to the median voter in $N$. 

Notice, it need not be the case that for any arbitrary space, a Condorcet winner exists, as shown by example in \cite{Olken_2014}. Moreover, when there is a cycle (or transitivity violation) in the social ordering, we say there exists a Condorcet cycle. It can be shown trivially that when there exists a Condorcet cycle, there is no Condorcet winner. Most of the foundational literature which we have referred to thus far considers orderings on a line, which as we mentioned, is indeed a median graph, but some more general connection is yet to be made between these concepts. We bring forward the notion of the Condorcet domain, or sets such that no plurality cycles exist. The work in \cite{puppe2019condorcet} brings together all these results by showing that every closed Condorcet domain is representable on a median graph, and every median graph corresponds to a closed Condorcet domain! Hence, the median graph is a good/reasonable space for social choice operate in -- and in his talk regarding this paper, Fain mentions that median graphs are the largest such graphs where the guarantee of the existence of a Condorcet winner is true \cite{Fain_2017}.

The median graph also has a beautiful connection with the Nash barganing problem discussed earlier. Let $u, v \in N$, with respective bliss points $p_u, p_v \in S$ and a particular alternative $a \in S$, which they deliberate over. In Lemma 1 \cite{fain2017sequential}, the authors prove that any median graph $G = (S, E)$ satisfies the property that the median of $(p_u, p_v, a)$ is exactly the maximizer of the Nash product, and an equilibrium of the Nash Bargaining problem.

\subsubsection{Distortion of the Algorithm}

The authors moved forward by finding bounds for distortion in this model. They find a mapping between any median graph $G$ into an isometric embedding $\phi(G)$ and were able to conclude that as the number of iterations increase, the distortion approaches 1.208, exponentially quickly.

The authors first prove a lemma that the uniqueness of the median of three points $t,u,v\in S$ is preserved by their isometric embeddings $\phi(t),\phi(u),\phi(v)$. For an arbitrary median graph $G=(S,E)$, the distortion of sequential deliberation on $G$ is at most that on $\phi(G)$. This is proven by that, in both cases of sequential deliberation and bargaining, the isometric embedding of the median $\phi(o^t)$ would be chosen. This lends to the fact the social cost is preserved in both $G$ and $Q$. This holds with the generalized median as well, so the distortion of sequential deliberation in $Q$ is an effective upper bound for deriving the main theorem.

The central observation for the distortion upper bound in the median graph is that sequential deliberation defines a Markov Chain on $Q$. As time goes to infinity, the expected distortion can be analyzed by the states of the Markov Chain's stationary distribution. It remains to solve for the correct transition probabilities. On a hypercube, this is made easy because the median of three points is the dimension-wise majority. By inspecting the dimension-wise bits of all agents, each dimension simply becomes a 2-state Markov Chain, with transition probabilities in terms of $f_k$, the fraction of agents whose bliss point's kth bit is $1$. Thus, the stationary probabilties can be derived in terms of $f_k$ as well. It remains to sum up the expected social cost via linearity of expectation. The final distortion ratio can be shown to be the fraction of a linear over quadratic polynomial, which can be upper bounded by $1.208$. 

This 2-state Markov Chain setup leads to a straightforward analysis of its convergence rate. The probability the present two agents have the same value in a dimension (i.e. the probability the 2-state chain "couples") decreases exponentially with $T$, hence deriving an upper guarantee on $T$ for this probability to become arbitrarily small. The final expression directly implies the theorem.

The implications of this theorem are crucial to their argument. The authors survey lower bounds of sequential deliberation for simpler classes. Mechanisms for median graphs to choose outcomes amongst agents' bliss points, $V_N$ produce a Distortion of at least 2. This is shown with the case of the k-star graph, when each agent's bliss point is a unique vertex on the periphery. 

Similarly, mechanisms to choose either an outcome or a median of three points must have distortion at least 1.316. The author show this by arguing that, with high probability, the number of bit-wise ones among any three points in $V_N$, then applying loose union bounds to bound the social cost of any median of three points. This lower bound proves the superiority of this framework of iterative sequential deliberation to one-shot deliberation mechanisms.

A similar approach is taken to capture the generalization of sequential deliberation, mechanisms constrained to pick outcomes on shortest paths between pairs in $V_N$. The authors show in this case, Distortion must be at least $9/8$, by arguing the number of ones that must exist on the shortest path between two agents, and deriving from it the expected social cost. This lower bound gives a sense how close the result is to the best pairwise Pareto-efficient frameworks can be.

Apart from these comparisons, there are inherent properties of sequential deliberation that capture its efficiency. First, the stationary distribution the authors used to show the theorem is in fact unique to the Markov Chain, or equivalently, that the Markov Chain is aperiodic and irreducible. Second, the outcome chosen by sequential deliberation is ex-post Pareto-efficient on a median graph. As the subsequent section elaborates further on, this means there are no alternatives that all agents $v\in N$ all prefer instead. It is remarkable how, in the context of the chosen model of sequential deliberation as a complex Markov Chain, this property is still preserved.

\subsubsection{Game Theoretic Properties of this Mechanism}

The chosen mechanism has many desirable properties, which the authors point out. The first of these properties is Pareto efficiency. On a median graph, the outcome of the algorithm, $\hat{a}$, is ex-post Pareto efficient. That is, there is no other $a$ where $d(a, v) \leq d(\hat{a}, v)$ $\forall v \in N$, (excluding the possibility of equality for all $v$). Intuitively, this is clearly useful as there is no other clear option point where all the agents can be made "better" from their bliss points simultaneously. 

Another useful game theoretic property is that of sub-game perfect Nash equlibrium (SPNE). Specifically, telling the truth for every agent is a SPNE. Recall that a SPNE means means that every subgame has equilibrium behavior. Here the authors structured their mechanism as a game. 

In the first step of the game, as a game of incomplete information, nature randomly picks two agents. Then, the two agents play the non-cooperative bargaining game; the outcome chosen in the game becomes the new disagreement alternative. The players have the ability to pick an arbitrary decision point in the feasible set $S$, and if they fail to agree or settle, then the previous disagreement alternative is set as the next decision point. The authors once again isometrically embed any such median graph $G$ into the hypercube, and use backward induction (a common technique to prove SPNE) in the game over the graph. Each player's utility corresponds to the objective of minimizing distance from their bliss point to the outcome of the game. 

\subsubsection{Extension to General Metric Spaces}

After extensive analysis in the case of median graphs, the authors zoom out for general metric spaces. For general metric spaces, there is a tight and pessimistic bound of 3 when the space of alternatives and bliss points lie in some metric. The tightness is due to observing the worst case scenario for Pareto efficient deliberation outcomes. A simple first step analysis of one step of deliberation yields an upper bound of $3OPT$, where $OPT$ is the minimum social cost of $a^*\in S$.

Another advantage of sequential deliberation is its advantage in terms of the distribution of outcomes it produces. The authors show the second moment of deliberation outcomes is bounded by a constant. By inequalities for the first and second moment, we can bind the chance of observing an outcome $\gamma$ times the optimal social cost by $O(1/\gamma^2)$. By contrast, random dictatorship has an unbounded second moment. This is shown when considering a graph of two nodes, with fraction $f$ agents on the first and $1-f$ on the other. As $f\rightarrow 0$, the squared social cost goes to zero slower than the optimal social cost, resulting in an unbounded second moment. The implication of the second moment is that the deliberation outcome will be robust to outlier agents, instead having a well-defined probability mass around the central consensus.

This interpretation is further extended by comparing the two methods in cases of nearly unanimous instances. As the Appendix shows, under the case of $\epsilon$-unanimity, where all but a small fraction of agents have the same bliss point, sequential deliberation outperforms random dictatorship on median graphs with a Distortion of $1+\epsilon$ to $2-\epsilon$. This easily follows from the linear over quadratic bound in terms of $f_k$ used for proving the theorem summarized earlier under the isometric embedding of the graph. With $\epsilon$-unanimity, the domain for $f_k$ is restricted to $[0,\epsilon]$ (given the optimal bliss point is at the $0$ vector. This improves the bound to $1+\epsilon$.

The pairwise deliberation scheme is also the meaningful deliberation scheme, particularly when expanded to the $N$-person deliberation scheme. The authors inspect the case of the line metric. Any disagreement outcome $o$ will be the final outcome due to agents' inability to come to consensus on an alternative from a Nash product of utility gain perspective. The authors wittily note how that confirms our intuition of deliberation breaking down in large groups to random outcomes. 

Thus, even though the value of the pairwise deliberation scheme only comes through in structured metric spaces, it exhibits, even on general metric spaces, the same assumptions and aforementioned properties for optimal Nash bargaining schemes schemes. The authors show a Nash Bargaining outcome would lie on the shortest path between agents.

%




%
\section{Simulations \& Further Study}
We build a general framework for evaluating the sequential deliberation model, with which we aim to test the robustness of the model when opinion dynamics come into play. This is partially motivated by a related open question in the paper, which suggests an exploration of the improvement in convergence when an agent's bliss point shifts towards the other agent and the decision alternative after each pairwise bargaining step. Primarily though, we study how sensitive the model is to the requirement that each bargaining step results in the Nash bargaining solution. To do so, we add a ``selfishness'' parameter $\lambda_a$ to each agent $a \in A$. We then implement two alternative bargaining functions in addition to the Nash bargaining function outlined in Equation \ref{eq:1}, Selfish Nash and Unselfish Nash.

In the Selfish Nash function, we randomly perturb the Nash bargaining outcome in the direction of the most selfish agent, weighted by the difference in selfishness between the pair of agents. This represents a bargaining paradigm in which we assume an imbalance in bargaining power that can influence the disagreement outcome away from the Nash equilibrium in a non-deterministic way, which we assume to be quite common in real-world bargaining.

The Unselfish Nash bargaining function finds the disagreement outcome according to the standard Nash bargaining function, but then has each agent randomly shift its bliss point towards the other agent's bliss point and the disagreement alternative, weighted by $1/\lambda$. In this sense, the willingness of each agent to shift its bliss point after the bargaining process is inversely proportional to its selfishness. This reflects an optimistic cooperative bargaining paradigm in which agents are particularly compromising in that they reevaluate their utilities after bargaining.

\subsection{Implementation Details}
Our framework consists of a decision space $S$ modelled by points along a line in the interval $[0,1)$, which is a simple example of a median graph. Let $\mathcal{B}(u,v,a)$ be the Nash optimal output given by finding $s \in S$ maximizing Equation \ref{eq:1}. Our disutility in this space is given by $d(u,v) = |u-v|$ for $u,v \in S$. We define 3 ``types'' or coalitions of bliss points in this space by sampling each agent's bliss point $b_a$ from one of 3 randomly placed normal distributions in the space, each of standard deviation 0.05. Mathematically, we have $\mu_1, \mu_2, \mu_3 \sim U[S]$ and $b_a \sim \mathcal{N}(\mu \sim U[\{\mu_1, \mu_2, \mu_3\}], 0.05)$. We also have $\lambda_a \sim \mathcal{N}(1,0.1)$. This setup allows for a large degree of random variety across simulations while still ensuring that bliss points are not spread out so uniformly that we end up with a degenerate version of the problem in which all outcomes have roughly equal social cost.  We outline our bargain functions below, with simplifications made for the sake of brevity.

\begin{algorithm}[h]
\SetAlgoLined
    \SetKwInOut{Input}{Input}
    \SetKwInOut{Output}{Output}
  \Input{agents $u, v \in A$ with $\lambda_u > \lambda_v$, alternative $a \in S$}
 Find Nash bargaining output $o =\mathcal{B}(u,v,a)$\\
 shiftDirection = $(b_u - o)/|b_u - o|$\\
 weight = $\lambda_u - \lambda_v$\\
 noise = $sample(U[0.9,1])$\\
 o' = o + weight $\times$ noise $\times$ shiftDirection

 \Output{shifted decision outcome $o'$}
 \caption{Selfish Nash Bargaining}
 \label{alg:1}
\end{algorithm}

\begin{algorithm}[h]
\SetAlgoLined
    \SetKwInOut{Input}{Input}
    \SetKwInOut{Output}{Output}
  \Input{agents $u, v$ , decision alternative $a$}
 Find Nash bargaining output $o =\mathcal{B}(u,v,a)$\\
 \For{ agent permutations $(x,y) \in\{(u, v), (v, u)\}$ }{
     destination = $((b_y - b_x) + (a - b_x)) / 2$
     shiftDirection = destination/|destination|
     weight = $1 / \lambda_x$\\
     noise = $sample(U[0.9,1])$\\
     $b_x$ = $b_x$ + weight $\times$ noise $\times$ shiftDirection
 }
 
 \Output{Decision outcome $o$}
 \caption{Unselfish Nash Bargaining}
 \label{alg:2}
\end{algorithm}

\begin{figure}
     \centering
     \begin{subfigure}[h]{\linewidth}
         \includegraphics[width=\textwidth]{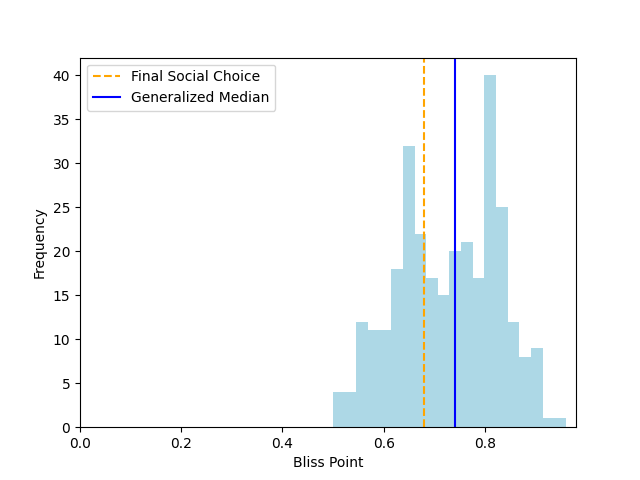}
         \caption{Perfect Nash Bargaining}
         \label{fig:p600}
     \end{subfigure}
     \vfill
     \begin{subfigure}[h]{\linewidth}
         \centering
         \includegraphics[width=\textwidth]{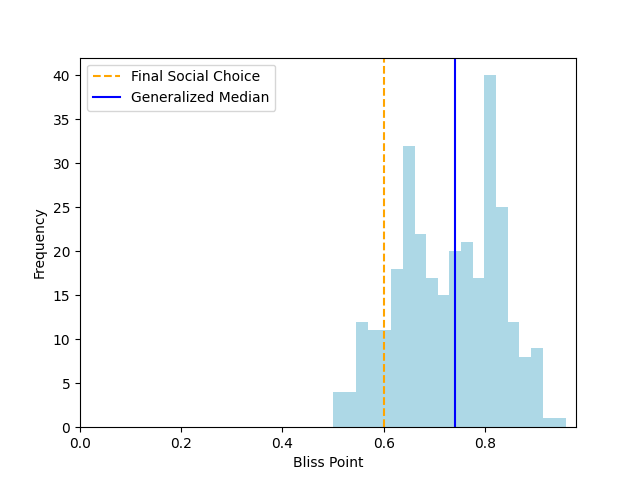}
         \caption{Selfish Bargaining}
         \label{fig:i600}
     \end{subfigure}
     \vfill
     \begin{subfigure}[h]{\linewidth}
         \centering
         \includegraphics[width=\textwidth]{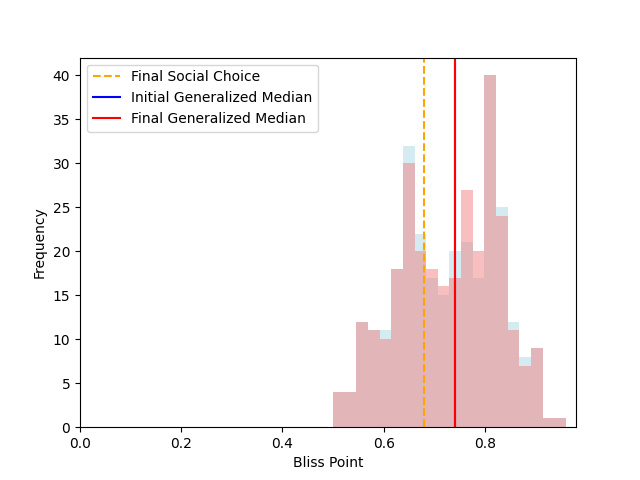}
         \caption{Unselfish Bargaining}
         \label{fig:b600}
     \end{subfigure}
        \caption{Distribution of agent bliss points with resulting social choice for all three bargaining functions on a single simulation, with common random seed. For \ref{fig:b600}, we show the initial (blue) and posterior (red) distributions after agent bliss points have shifted.}
        \label{fig:1}
\end{figure}

The rest of the sequential deliberation simulation algorithm follows Figure 1 in \cite{fain2017sequential}. For our simulation, we set the number of alternatives and number of agents as $|S| = 50$ and $|A| = 300$. We limit the number of deliberation steps to 10.  We run 1000 simulations with each of the three bargaining paradigms.

\subsection{Simulation Results and Discussion}

\begin{figure}
     \centering
     \begin{subfigure}[h]{\linewidth}
         \includegraphics[width=\textwidth]{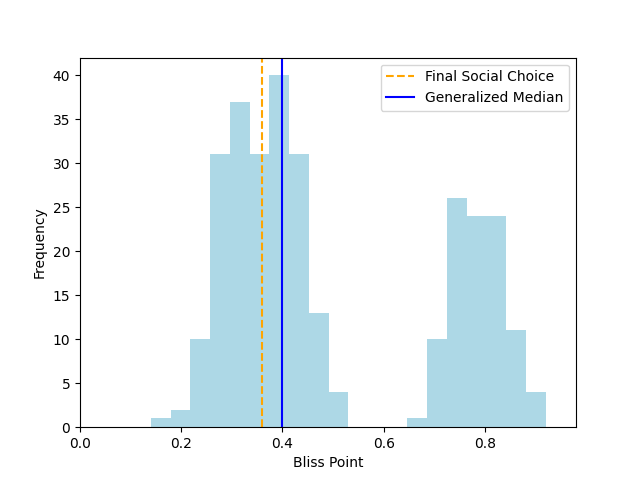}
         \caption{Perfect Nash Bargaining}
         \label{fig:p600b}
     \end{subfigure}
     \vfill
     \begin{subfigure}[h]{\linewidth}
         \centering
         \includegraphics[width=\textwidth]{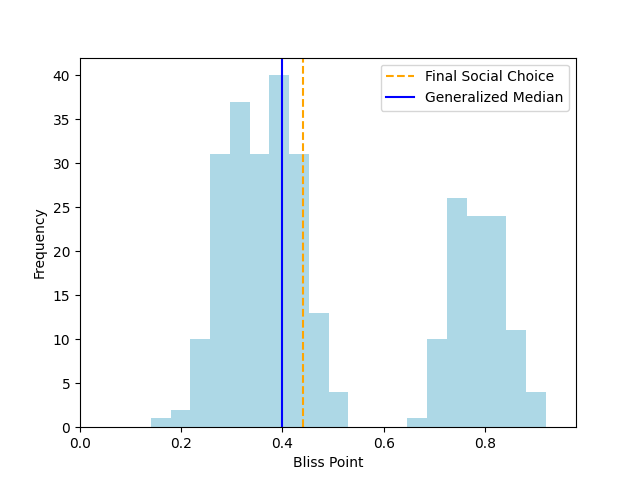}
         \caption{Selfish Bargaining}
         \label{fig:i600b}
     \end{subfigure}
     \vfill
     \begin{subfigure}[h]{\linewidth}
         \centering
         \includegraphics[width=\textwidth]{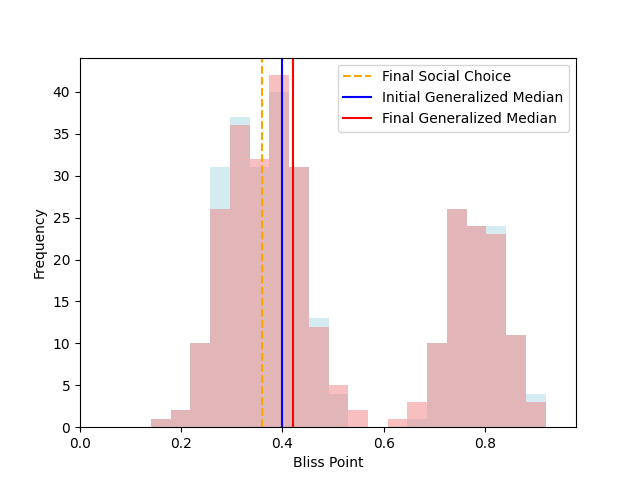}
         \caption{Unselfish Bargaining}
         \label{fig:b600b}
     \end{subfigure}
        \caption{Another distribution of agent bliss points with resulting social choice for all three bargaining functions on a single simulation, with common random seed. For \ref{fig:b600b}, we show the initial (blue) and posterior distributions (red) after agent bliss points have shifted.}
        \label{fig:2}
\end{figure}

Figures \ref{fig:1} and \ref{fig:2} show two examples of our simulation. In \ref{fig:i600}, we observe that the selfish bargaining function arrives at a social choice that is much further from the generalized median than in the perfect Nash scenario. This effect was consistent across simulations, and reflects the social effect of those with higher bargaining power influencing deliberation towards their own bliss points and away from the Nash solution. In \ref{fig:b600}, note that the distribution of bliss points after deliberation is clustered closer to the generalized median, reflecting the intended effect of agents' opinions shifting towards one another through the deliberation process. This process is stochastic and not necessarily symmetric about the generalized median, and as a result, the optimal social choice can actually change over the course of deliberation as shown in \ref{fig:b600b}. 

\subsubsection{Distortion}
\begin{figure}
     \centering
     \begin{subfigure}[h]{\linewidth}
         \includegraphics[width=\textwidth]{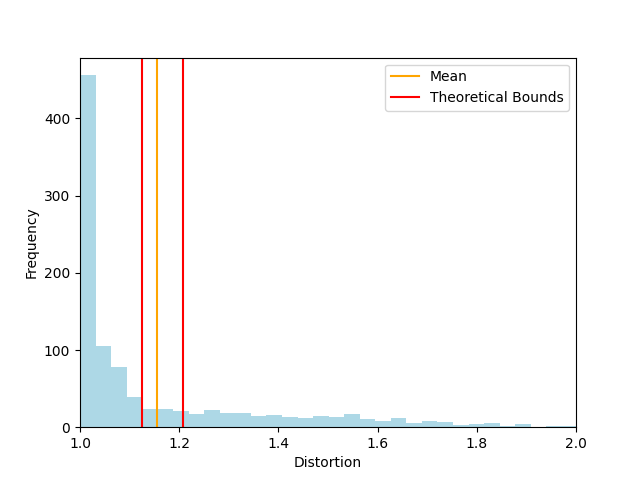}
         \caption{Perfect Nash Bargaining}
         \label{fig:3a}
     \end{subfigure}
     \vfill
     \begin{subfigure}[h]{\linewidth}
         \centering
         \includegraphics[width=\textwidth]{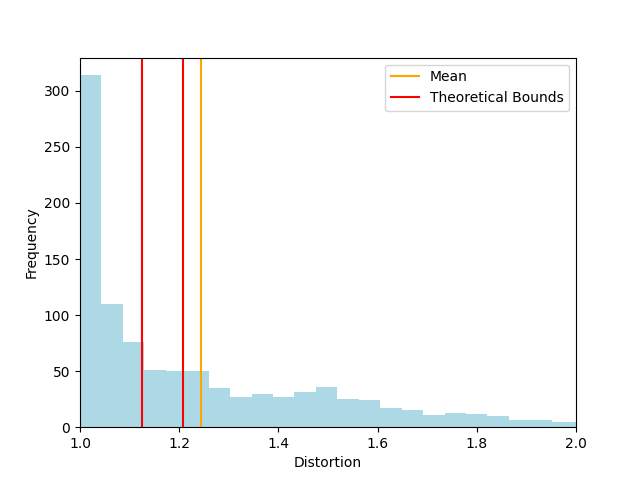}
         \caption{Selfish Bargaining}
         \label{fig:3b}
     \end{subfigure}
     \vfill
     \begin{subfigure}[h]{\linewidth}
         \centering
         \includegraphics[width=\textwidth]{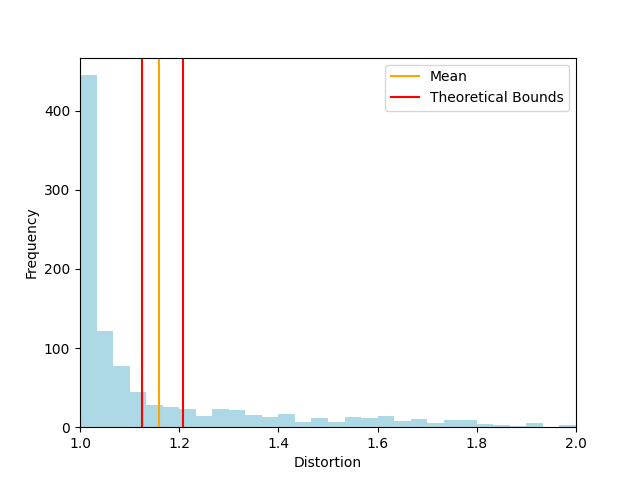}
         \caption{Unselfish Bargaining}
         \label{fig:3c}
     \end{subfigure}
        \caption{Distribution of Distortion for final social choice after pairwise sequential deliberation for each of the three bargaining functions. The theoretical bounds reflect the lower bound of 1.125 and the upper bound of 1.208 proven for Nash bargaining in \cite{fain2017sequential}.}
        \label{fig:3}
\end{figure}

From \cite{fain2017sequential}, we know that the upper bound distortion for infinite-horizon sequential deliberation with Nash bargaining is 1.208 and the lower bound for pairwise Pareto-efficient deliberation schemes is 1.125. We use these as baselines in Figure \ref{fig:3}, where we show the distribution of final Distortions for all 1000 simulations in each bargaining scheme. With a mean Distortion of 1.16, our simulated Nash bargaining scheme (Figure \ref{fig:3a}) does indeed fall within these bounds.

Interestingly, our selfish bargaining function (Figure \ref{fig:3b}) - which violates the assumption of arriving at a Nash-optimal alternative in each step - shows a much longer tail in its distribution which results in a mean Distortion of 1.24. This violates the 1.208 ideal upper bound for the Nash-optimal bargaining scheme and indicates that the Nash-optimality requirement is strict in practice and that the efficacy of the sequential deliberation framework is quite sensitive to the optimality of the bargaining function. Therefore, the pairwise sequential deliberation scheme proposed in \cite{fain2017sequential} is not necessarily robust in situations where opinion dynamics are such that the bargaining outcome in each step can deviate from the Nash bargaining solution.

In Figure \ref{fig:3c}, we note that the unselfish bargaining scheme results in a distribution that is remarkably similar to the perfect Nash bargaining scheme. This is likely due to the fact that in a simple median graph such as our decision space and with only at most 20 agents sampled from over 300, converging agent utilities after bargaining has too small an effect on the global distribution of bliss points to significantly affect the Distortion. Indeed, we found that the mean Distortion in this case is 1.17, slightly worse than the Nash bargaining function. Note that one of the strengths of pairwise deliberation is being able to arrive at a near-optimal outcome without sampling a large proportion of agents from the population. This effect counteracts the gain in convergence from agents' utilities shifting, as few bliss points will have shifted relative to the overall distribution of agent bliss points. Additionally, the stochastic nature of this scheme shifts the generalized median throughout the deliberation process, and we suspect that this may have increased the mean Distortion. However, it is important to note that this may not be true for metric spaces that require sampling a larger proportion of the population, in which case the shifting utilities may noticeably improve the final Distortion.

\subsubsection{Convergence}

\begin{figure*}[t]
         \includegraphics[width=\textwidth]{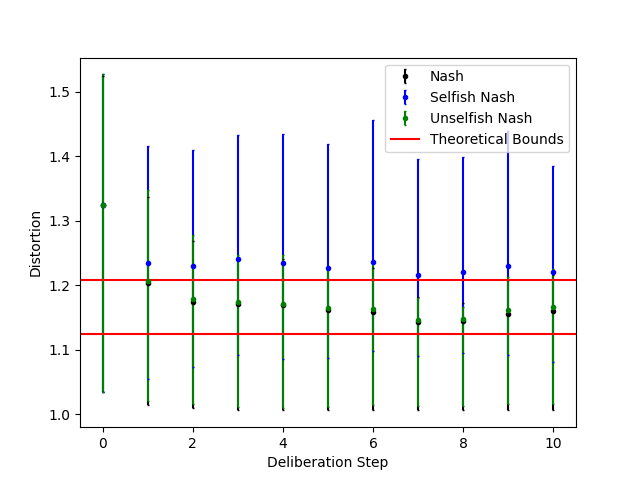}
        \caption{Distribution of Distortion for final social choice after pairwise sequential deliberation for each of the three bargaining functions. The theoretical bounds reflect the lower bound of 1.125 and the upper bound of 1.208 proven for Nash bargaining in \cite{fain2017sequential}.}
        \label{fig:4}
\end{figure*}

We also study the convergence behaviour of the framework. To do so, we plot the mean Distortion across all 1000 simulations along with the first and third quartile Distortions for each of the three bargaining schemes in Figure \ref{fig:4}. The rapid convergence of this framework is evident - with Nash or Unselfish bargaining, the mean distortion falls below 1.2 in 2 steps. The third quartile for these schemes decreases steadily as the number of deliberation steps increases, with a final third quartile Distortion just over 1.21 at step 10. Interestingly, the first quartile reveals that Nash bargaining tends to find the optimal social choice slightly more often than Unselfish bargaining. In a space of 300 agents, convergence to a nearly optimal outcome in such a small number of pairwise bargaining steps is rather remarkable.

Selfish bargaining stabilizes within one step, with neither the mean nor the quartiles showing significant improvement after the first step. This suggests that there is a lower bound for the Distortion under the Selfish scheme. This lower bound - around 1.22 in our case - is likely dictated by the variance in selfishness across agents and the magnitude of the expected weight parameter in the Selfish bargaining algorithm. We also note that Selfish bargaining has the largest inter-quartile range, primarily because the Selfish scheme does not result in any decrease of the third quartile Distortion as deliberation progresses beyond step 1. This again indicates that the sequential deliberation framework is significantly compromised without a Nash-optimal bargaining result.

\subsection{Conclusion \& Future Works}

The results presented in \cite{fain2017sequential} are a promising technique for arriving at a conclusion amongst a group of agents where the decision space is potentially unenumerable. Especially, when comparing against any deterministic algorithm, the distortion is almost always a significant improvement. However, we strongly believe that this mechanism operates via strong repeated assumptions of bargaining according to Nash bargaining, which may not always translate in many environments. 

In situations like this, mechanism designers often explain the importance of experimentation to test the validity of these techniques; despite the complexity of experimental design for such a mechanism, (hidden preferences, defining the alternative space, etc.) we feel that formal experimental design would be a highly interesting idea, beyond the experiments which have already been conducted on Nash's work. This is because, in models with nearly 20-30 iterations of an assumption can look very different from 1 iteration -- especially as agents see how games play out. repeated able to Throughout this process of simulation, 

\subsubsection{Simulation Framework}
Our simulation framework is general and easily extendable to other metric spaces and bargaining functions. We would like to run similar experiments on more complex metric spaces. We believe that the metric space is a delicate component of the framework, much like the bargaining function, and hypothesize that as the complexity of the graph grows, the results deviate further from the optimum.

\subsubsection{Orderings}

Many times, an agent is interested in learning the ordering of the top-$k$ elements in a decision space. We were interested in learning how rankings would operate in this space. For example, given $S$ lives in the discrete line graph $[0, 9]$ (that is, it contains exactly $10$ vertices, each connected to the adjacent vertex). One could augment this decision space to $S'$ with $10!$ states, each corresponding to a distinct ordering of the states in $S$. Some sort of ordering can be introduced to this graph, whether it be lexicographical \cite{golenko1991metrics} or transposition distance. 

In the case of lexicographical distance, we believe that asking all agents to pick a bliss point corresponding to their ordering is potentially equivalent to running $k$ independent cycles of the mechanism, and for each iteration $i$, picking the bliss point corresponding to their $i^{th}$ most preferred outcome. 

In future experimentation and exploration, we seek to investigate the validity of these claims, and also explore how orderings can realistically be embedded from an original graph $S$ to a new graph $S'$, while maintaining the original structure of the old graph.

\vfill\null 
\bibliography{ref}

\end{document}